\documentclass[twocolumn,trackchanges]{aastex701}
\usepackage{amsmath}
\usepackage{CJK}
\linenumbers

\begin{document}
\begin{CJK*}{UTF8}{gbsn}

\title{Quasi-periodic Eruptions from Stellar-mass Black Holes Impacting Accretion Disks in Galactic Nuclei}

\author[orcid=0000-0001-8916-2775]{Kun Liu (刘坤）}
\affiliation{School of Physics and Astronomy, Sun Yat-sen University, Zhuhai 519082, China}
\email{liuk89@mail2.sysu.edu.cn} 

\author[orcid=0000-0002-9442-137X]{Shang-Fei Liu (刘尚飞)} 
\affiliation{School of Physics and Astronomy, Sun Yat-sen University, Zhuhai 519082, China}
\affiliation{CSST Science Center for the Guangdong-HongKong-Macau Great Bay Area, Sun Yat-sen University, Zhuhai 519082, China}
\email[show]{liushangfei@mail.sysu.edu.cn}

\author[orcid=0000-0001-9608-009X]{Zhen Pan (潘震)}
\affiliation{Tsung-Dao Lee Institute, Shanghai Jiao-Tong University, 520 Shengrong Road, Shanghai 201210, China}
\affiliation{School of Physics \& Astronomy, Shanghai Jiao-Tong University, 800 Dongchuan Road, Shanghai 200240, China}
\email{zhpan@sjtu.edu.cn}

\author[orcid=0000-0001-6858-1006]{Hongping Deng (邓洪平)}
\affiliation{Shanghai Astronomical Observatory, Chinese Academy of Sciences, 80 Nandan Road, Shanghai 200030, China}
\email{hpdeng353@shao.ac.cn}

\author[orcid=0000-0001-5012-2362]{Rongfeng Shen (申荣锋)}
\affiliation{School of Physics and Astronomy, Sun Yat-sen University, Zhuhai 519082, China}
\affiliation{CSST Science Center for the Guangdong-HongKong-Macau Great Bay Area, Sun Yat-sen University, Zhuhai 519082, China}
\email{shenrf3@mail.sysu.edu.cn}

\author[orcid=0000-0003-0454-7890]{Cong Yu (余聪)}
\affiliation{School of Physics and Astronomy, Sun Yat-sen University, Zhuhai 519082, China}
\affiliation{CSST Science Center for the Guangdong-HongKong-Macau Great Bay Area, Sun Yat-sen University, Zhuhai 519082, China}
\email{yucong@mail.sysu.edu.cn}
\correspondingauthor{Shang-Fei Liu}

\begin{abstract}

We investigate the origins of quasi-periodic eruptions (QPEs) in galactic nuclei using global three-dimensional meshless finite-mass (MFM) simulations. By modeling stellar and black-hole impactors traversing accretion disks under various inclinations and surface densities, we evaluate their consistency with the observed properties of QPEs. Stellar impacts produce highly asymmetric bipolar ejecta with forward outbursts dominating by over an order of magnitude in energy and luminosity due to the star blocking downstream flow and creating a low-density wake. This shock-compression mechanism often renders backward events unobservable, implying one detectable burst per orbit, and challenging the standard assumption of two bursts. It also fails to explain alternating long--short recurrence patterns and places several sources near or within twice the tidal disruption radius for solar-mass stars, raising severe stability concerns. Whereas a stellar-mass black hole (sBH) gravitationally focuses and heats disk gas over an effective interaction scale that extends beyond its Bondi radius $R_{\rm B}$ and is naturally bounded by its Hill radius $R_{\rm H}$ during an impact, yielding nearly symmetric ejecta with mild contrasts. This gravitational-drag mechanism generates higher energy budgets at low inclinations due to enhanced mass accumulation. We suggest an ad hoc effective interaction radius $  R_{\rm eff} \simeq 0.5\, R_{\rm B}^{1/3} R_{\rm H}^{2/3}  $ to quantify this trend. Incorporating this effective radius substantially increases the energy that sBH-disk collisions can produce compared to previous Bondi-only estimates, improving the viability of stellar-mass black holes as the impactors for a wide range of observed QPE energies and properties.

\end{abstract}

\keywords{\uat{High Energy astrophysics}{739} --- \uat{X-ray transient sources}{1852} --- \uat{Black holes}{162} --- \uat{Active galactic nuclei}{16} ---\uat{Hydrodynamics}{1963}  --- \uat{Tidal disruption}{1696}}

\section{Introduction} 
\label{sec:intro}

\begin{figure*}[ht!]
\plotone{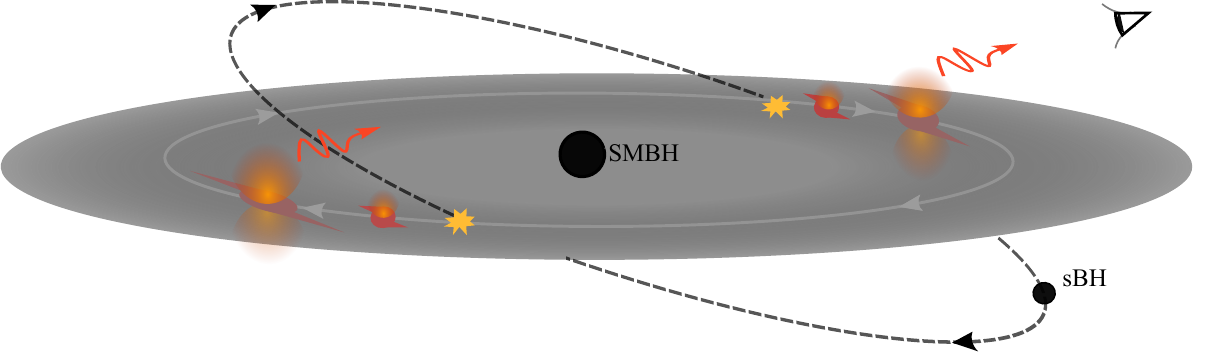}
\caption{Schematic illustration of an extreme mass-ratio inspirals (EMRIs) system, where a stellar-mass black hole (sBH) moves along a low-inclination orbit and penetrates an optically thick accretion disk with a relative velocity below the local Keplerian speed. During each orbital period, the sBH impacts the disk twice, exciting nearly symmetric ejecta in the vertical directions. The ejecta subsequently expand and cool, evolving into optically thin bubbles through which radiation can escape. Meanwhile, strong gravitational perturbations induce local spiral-shaped density waves in the disk material.
}
\label{fig:QPE}
\end{figure*}

Quasi-periodic eruptions (QPEs) represent a puzzling new class of soft X-ray transients discovered in the past decade, characterized by recurring flares superimposed on a quiescent continuum of active galactic nuclei (AGNs). QPE durations can be as long as $10^{5}$ seconds, with recurrence intervals from hours to days and flux increases of $\sim$10--100 times in the soft X-ray band (0.1--0.2 keV) \citep{giustini2020A&A...636L...2G, chashkina2024A&A...691A.313C}. The first QPEs were discovered by the XMM-Newton satellite in GSN 069 \citep{miniutti2019Natur.573..381M} and RX J1301.9+2747 \citep{Sun2013,giustini2020A&A...636L...2G}, followed by eROSITA detections of eRO-QPE1-5 \citep{arcodia2021Natur.592..704A,arcodia2024A&A...684A..64A,arcodia2025ApJ...989...13A}. Recent additions include Ansky, AT2022upj and J2344 in 2025 and 2026 \citep{ansky2025NatAs...9..895H, Chskrsborty2025ApJ...983L..39C, baldini2026A&A...706L..15B}, highlighting their diversity and potential prevalence in time-domain surveys.

QPEs often link to tidal disruption events (TDEs); for instance, they appear $\mathcal{O}(1)$ year post-ignition in AT2019qiz \citep{nicholl2024Natur.634..804N}
, XMMSL J0249 \citep{chakraborty2021ApJ...921L..40C} and among several hosts with TDE signatures \citep{Jiang:2025jbd}. \cite{Chakraborty:2025ntn} further estimate that $\sim 10\%$ optical TDEs result in QPEs within 5 yr post-disruption.
In parallel, a newly activated AGN \citep{ansky2025NatAs...9..895H} or a featureless TDE \citep{Zhu2025ApJ...994L..16Z}, host the most energetic QPEs, Ansky, detected so far, exhibiting unusually powerful and puzzling period modulations \citep{Chakraborty:2026opq}. Recent observations indicate that QPEs show a preference for a subset of TDEs in galaxies hosting an extending emission line region \citep{Wevers2022,Wevers2024,Xiong:2025gez} and/or a parsec scale dust torus \citep{Wu:2025vgt}.

Unlike the stochastic variability commonly observed in AGN, QPEs commonly exhibit regular patterns, including alternating long -- short recurrence times ($T_{\mathrm{long}},\,T_{\mathrm{short}}$) or strong -- weak amplitudes \citep{miniutti2019Natur.573..381M,giustini2020A&A...636L...2G}. 
\cite{zhou2024PhRvD,zhou2024PhRvD.110h3019Z} further pointed out that the sum of two consecutive intervals $T_{\rm long}+T_{\rm short}$ is often constant, though both $T_{\rm long}$ and $T_{\rm short}$ show clear variations. The eruptions appear as soft X-ray outbursts at 0.1--0.2 keV against a cooler quiescent continuum (0.05--0.08 keV), with flare energies of $10^{45}$ -- $10^{48}\,\mathrm{erg}$, well below those of tidal disruption events but vastly larger than stellar flares, thereby constraining the underlying energy reservoir.
Meanwhile, their hosts typically harbor low-mass supermassive black holes (SMBHs; $M_{\mathrm{BH}} \sim 10^{5-7}M_\odot$ \citep{arcodia2024A&A...684A..64A}), suggesting orbital dynamics in compact regions.

The physical origin of QPEs remains the subject of active debate.  
A family of models invokes a star on a bound orbit around a supermassive black hole that supplies material to the central engine. In the repeating partial tidal disruption (rpTDE) variant, the star undergoes repeated partial stripping near pericenter, launching short-lived X-ray outbursts \citep{evans2023NatAs...7.1368E, pasham2024arXiv241105948P}. A closely related picture involves steady mass transfer from the orbiting body onto the central black hole, for example through Roche-lobe overflow (RLOF) in stars \citep{zalamea2010MNRAS.409L..25Z,king2020MNRAS.493L.120K,zhao2022A&A...661A..55Z,krolik2022ApJ...941...24K,metzger2022ApJ...926..101M,wang2022ApJ...933..225W,lu2023MNRAS.524.6247L, linial2023ApJ...945...86L}. Although both pictures can produce periodic flares under favorable conditions, they struggle to sustain stable outbursts over hundreds of cycles and offer no natural mechanism for the observed alternating long--short recurrence pattern. Moreover, the formation rate of such tight stellar orbits (hours to days) — whether through the loss-cone channel, the Hills mechanism, or Roche-lobe capture — remains orders of magnitude too low to account for the observed QPE population \citep{Pan:2026awf}.

The disk instability model attributes QPE outbursts to intrinsic processes within the accretion disk itself—thermal instabilities, magnetorotational instabilities (MRI), or other spontaneous perturbations \citep{raj2021ApJ...909...82R, pan2022ApJ...928L..18P, pan2023ApJ...952...32P, Kaur2023MNRAS.524.1269K, sniegowska2023A&A...672A..19S, Deng2025}. With suitable inner-boundary torque and stress prescriptions, this framework can successfully reproduce several observed light-curve and spectral properties of GSN 069 \citep{pan2022ApJ...928L..18P}. However, because these instabilities are generated internally, the model lacks a robust external clock capable of enforcing the strict repeating periodicity observed in most QPE sources. Consequently, it struggles to account for the characteristic alternating long--short recurrence intervals and strong--weak amplitude pattern unless additional modulating physics is introduced.

The current most popular explanation invokes an extreme mass ratio inspiral (EMRI) that crosses the accretion disk twice per orbital period, thereby producing two eruptions per orbit \citep[see, eg. Fig. \ref{fig:QPE}; ][]{dai2010MNRAS.402.1614D, xian2021ApJ...921L..32X,sukova2021ApJ...917...43S,franchini2023A&A...675A.100F,linial2023ApJ,tagawa2023MNRAS.526...69T,linial2024ApJ...963L...1L,zhou2024PhRvD,zhou2024PhRvD.110h3019Z,zhou2025ApJ...985..242Z,zhou2025arXiv250411078Z, vurm2025ApJ}. The disk itself is naturally supplied by a recent tidal disruption event, which simultaneously accounts for the TDE-QPE association\citep{linial2023ApJ, nicholl2024Natur.634..804N}. A mildly eccentric EMRI orbit or differences in the impact angle then naturally gives rise to the alternating long--short recurrence intervals and strong--weak flare amplitudes that are characteristic of many sources.

However, it has been claimed that the long--short recurrence pattern in GSN 069 and eRO-QPE2 is accompanied by correlated (rather than anti-correlated) timing variations between odd and even flares \citep{arcodia2024A&A...690A..80A, miniutti2025A&A...693A.179M, arcodia_even_2026}. As pointed out by \citet{2026arXiv260529382Z}, the constant $T_{\rm long}+T_{\rm short}$ in recurrence times of GSN 069 indicates that timing variations between odd and even flares are anti-correlated. The opposite claim is likely due to a mismatch in the cycle number of eruptions in the timing analysis.  For eRO-QPE2, \citet{2026arXiv260529382Z} found that the QPE timing, including the correlated timing variations, is  well modeled by a near-circular EMRI + a precessing disk, where the EMRI apsidal precession leaves little imprint on the QPE recurrence times, while the disk precession contributes to the correlated timing variations.

Whether two flares per orbit are seen in all QPEs remains under active debate. Here, we primarily focus on the single-impact physics, with particular emphasis on ejecta asymmetry, energy generation mechanisms and resulting budgets(shock compression versus gravitational drag), and the role of the Hill radius. A clear understanding of these single-impact processes is essential for correctly interpreting long-term orbital evolution and the more sophisticated timing mechanisms now being revealed by high-cadence observations.

The question of whether the secondary companion is a star or a compact object remains open. 
\citet{linial2023ApJ} argue strongly in favor of a main-sequence star, pointing out that its physical radius ($\sim R_\odot$) is orders of magnitude larger than the Bondi radius of a stellar-mass black hole ($\sim 0.02 R_\odot$ for a perpendicular passage of a $100\, M_\odot$ black hole). Consequently, a stellar EMRI shocks a much larger column of disk gas per passage, whereas a stellar-mass black hole would simply produce too little shocked material to explain the observed flare luminosities unless the secondary is an intermediate-mass black hole.

Local hydrodynamic simulations that model a stellar EMRI crossing a stationary patch of the accretion disk have been used to test this picture. \citet{yao2025ApJ...978...91Y} show that repeated passages drive significant mass loss and expansion of the stellar envelope, limiting the star's lifetime to at most $  \sim 1000\,\mathrm{yr} $. The authors therefore propose that the observed QPEs are powered primarily by collisions between the expanding stellar debris stream and the disk, rather than by the intact star itself. 
\citet{huang2025ApJ...993..186H} performed 2D multi-group radiation-hydrodynamic simulations of a solar-type star colliding with a thin disk and showed that the interaction produces asymmetric bipolar ejecta, with prominent forward breakout emission leading to sharply peaked light curves and large luminosity contrasts between the two sides. The resulting flares reach peak bolometric luminosities $  \gtrsim 10^{43}\,\mathrm{erg\,s^{-1}}  $, have durations of sub-hour to a few hours, and exhibit SEDs that peak in the extreme-UV (20--50 eV) with soft X-ray luminosities $  \nu L_\nu \gtrsim 10^{42}\,\mathrm{erg\,s^{-1}}  $, broadly consistent with short-period QPEs. However, both studies highlight persistent challenges: strong ejecta asymmetry, relatively short flare durations, and the difficulty of sustaining long-term stable outbursts.

In reality, the Hill radius of a stellar EMRI is only $\sim$ 2--3 times larger than its physical radius. Local simulations \citep{yao2025ApJ...978...91Y, huang2025ApJ...993..186H} that neglect the central SMBH's gravity effectively allow the Hill radius to extend to the boundary of their computational domain. If this domain is significantly larger than the true Hill radius, the gas reservoir available for perturbation could be overestimated, making a stellar EMRI appear more effective at powering QPEs than it would be in a realistic global potential. In the opposite direction, semi-analytical calculations such as \citep{linial2023ApJ} ignore the Hill radius entirely. In striking contrast, for a 100 $M_\odot$ black hole the Hill radius reaches $R_\textrm{H} \sim 10 R_\odot$, which is orders of magnitude larger than the Bondi radius $\sim 0.02\, R_\odot$ used by them to estimate the shocked mass. As a result, a substantial reservoir of gas lying between the Bondi and Hill spheres can be gravitationally perturbed, heated, and later ejected, strongly affecting both the total energy budget and the symmetry of the two-sided flares.

To summarize, global three-dimensional hydrodynamic simulations that self-consistently include the full gravitational field of the SMBH are essential to properly resolve the relative contributions of the Bondi and Hill scales, quantify the true shocked-gas mass and energy budget, and assess whether stellar-mass black holes can power the observed QPEs. While such global simulations have recently been performed for more massive perturbers \citep{ressler2024ApJ...967...70R, ressler2025ApJ...993L..22R, 2025arXiv250619900D}, which can excite spiral density waves and drive longer-term disk evolution, simulations in the low-mass-ratio EMRI regime relevant to QPEs have not yet been reported in the literature.

This paper is organized as follows. In Section~\ref{sec:model}, we analyze the theoretical framework of the EMRI+disk model and estimate relevant timescales and energy budgets. In Section~\ref{sec:sim}, we then describe the setup of our global numerical simulations. Results of star-disk and black hole–disk collisions are presented in Section~\ref{sec:results}. In Section~\ref{sec:discussion}, we offer a detailed discussion and further analysis of the EMRI+disk collision model.

\section{A Quick Recap of EMRI+disk Models} \label{sec:model}

The EMRI+disk model has gained increasing support from both observational evidence and theoretical analysis. In this model, the energy budget of eruptions comes from EMRI-disk collisions, and two eruptions are produced as the EMRI crosses the disk twice per orbital period.
The orbital eccentricity naturally gives rise to alternating short-long intervals and strong -- weak outbursts. The  Schwarzschild (and Lense-Thirring) precessions of the EMRI orbit cause the intersection point between the orbit and the disk to vary, leading to a changing impact phase from one orbit to the next \citep{dai2010MNRAS.402.1614D,xian2021ApJ...921L..32X,zhou2025arXiv250411078Z}. These relativistic precession effects lead to a modulation between two outburst intervals, $T_{\mathrm{short}}$ and $T_{\mathrm{long}}$, while keeping the sum $T_{\mathrm{short}} + T_{\mathrm{long}}$ approximately a constant \citep{zhou2024PhRvD,zhou2025ApJ...985..242Z}. This behavior is broadly consistent with observations of GSN 069 and several other QPE sources. Therefore, the EMRI+disk model not only offers a compelling explanation for the energy budget and burst characteristics, but also provides a clear and readily testable theoretical framework for the origin of QPEs.

In EMRI systems with an sBH as the secondary, passage through the surrounding disk gas can lead to accretion of disk material and the formation of a transient accretion flow, producing radiation powered by accretion. This process can trigger periodic X-ray outbursts, with the luminosity constrained by the Eddington limit
\begin{equation}
    L_{\text{Edd}} \approx 1.26 \times 10^{38} \left( \frac{M_{\mathrm{BH}}}{M_\odot} \right) \, \mathrm{erg/s}.
\end{equation}
In most QPE events, the observed burst luminosities often exceed $10^{42} \, \mathrm{erg/s}$. If this eruption is attributed to Eddington-limited accretion, the required black hole mass must exceed $10^4 M_\odot$. On the other hand, if the black hole shocks materials within its Bondi radius
\begin{equation}
    R_{\mathrm{B}} = \frac{2GM_{\mathrm{BH}}}{v_{\mathrm{rel}}^2},
\end{equation}
a black hole with a mass of $10^4 M_\odot$ would still be required to shock a sufficient amount of gas, even with a relatively large impact velocity of 0.1c \citep{vurm2025ApJ}.  However, the rarity of intermediate-mass black holes (IMBHs), with no firm detection yet, renders black hole impacts less favourable than star-disk collisions. 

Recent local general-relativistic axisymmetric 2D simulations have examined the collisions of more massive BHs ($\sim 10^4 M_\odot$) on accretion disks with an emphasis on the radiative consequences of post-collision accretion \citep{lam2025PhRvD.112h3006L}. Global general relativistic magnetohydrodynamic (GRMHD) simulations have also been conducted to model SMBH-disk collisions in the context of the SMBH binary candidate OJ 287 \citep{ressler2024ApJ...967...70R,ressler2025ApJ...993L..22R}. However, the parameter space and physical regime explored in these studies differ significantly from the sBH collisions considered here. In addition, the effects of the impact inclination are often not fully considered, which is directly relevant to the size of the Bondi radius. Previously, this aspect has been specifically examined by \citet{franchini2023A&A...675A.100F, tagawa2023MNRAS.526...69T}.

Here, we highlight the significant impact of the inclination angle $i$ and the disk motion on the relative velocity between the impactor and the disk, thereby strongly influencing various physical parameters such as the energy budget. For a Keplerian disk, the relative velocity can be written as
\begin{equation}
    v_{\mathrm{rel}} = 2v_{\mathrm{k}}\sin\left(\frac{i}{2}\right), 
\end{equation}
with $v_k$ being the Keplerian velocity. The relative velocity is $v_{\mathrm{k}} \sin i$ if the disk is taken as static, which underestimates the true relative velocity, especially at high inclinations. Moreover, the path length traversed by the impactor within the disk is:  

\begin{equation}
    \Delta L = 2H/\cos(i/2),
\end{equation}
where $H$ is the vertical scale height of the disk.

In the stellar EMRI+disk model, a star periodically crosses a TDE disk around a SMBH \citep{linial2023ApJ}, with nearly vertical and circular orbits commonly adopted for illustrative purposes and energy estimates. During each disk crossing, the star drives shocks that locally heat the gas, which then expands adiabatically and may emit soft X-rays as blackbody radiation. The mass of shocked gas per collision is expressed as \citep{linial2023ApJ}, 
\begin{equation}
M_{\mathrm{sh}} = \pi R_{\mathrm{eff}}^{2} \Sigma/\cos(i/2),
\end{equation}
where $\Sigma$ denotes the local surface density of the accretion disk, and the effective radius $R_{\mathrm{eff}}$ is the actual radius over which the disk gas is shocked, usually corresponding to the stellar radius $R_*$ in the star-disk model and to the Bondi radius $R_{\mathrm{B}}$ in the sBH-disk model. The factor of $1/\cos(i/2)$ is included because the simplified 2D disk model applies only to collisions perpendicular to a stationary disk, whereas in our case the path length traversed by the impactor varies with the orbital inclination. Assuming the shocked gas is accelerated to the same velocity as the impactor, and for convenience fixing the orbital inclination at $i = \pi/10$, the energy budget of a QPE burst can be approximated by  
\begin{equation}
\begin{aligned}
    E_{\mathrm{sh}} &=  \frac{1}{2}M_{\mathrm{sh}}v_{\mathrm{rel}}^{2} \\
&\approx 4.3\times10^{45}\mathrm{ergs}\left(\frac{\Sigma}{10^{6}\mathrm{g\,cm^{-2}}}\right) \left(\frac{R_{\mathrm{eff}}}{R_{\odot}}\right)^{2}\left(\frac{v_{\mathrm{rel}}}{0.025c}\right)^{2}\ .
\end{aligned}
\end{equation}
Here $R_{\mathrm{eff}}$ denotes the effective interaction radius introduced above. In the star-disk model, $R_{\mathrm{eff}} = R_*$ is independent of the relative velocity, leading to $E_{\mathrm{sh,st}} \propto v_{\mathrm{rel}}^{2}$. In contrast, in the sBH-disk model, the Bondi radius scales as $R_{\mathrm{B}}\propto v_{\mathrm{rel}}^{-2}$, which results in $E_{\mathrm{sh,bh}} \propto v_{\mathrm{rel}}^{-2}$. The kinetic energy deposited in the shocked disk material represents only an upper bound on the radiated energy, because part of the energy may be lost through adiabatic expansion or emitted outside the observed bands, which may also affect the asymmetry of the observed radiation\citep{huang2025ApJ...993..186H,jankovi2026arXiv260202656J}.

\begin{figure}[ht!]
\centering
\includegraphics[width=\columnwidth]{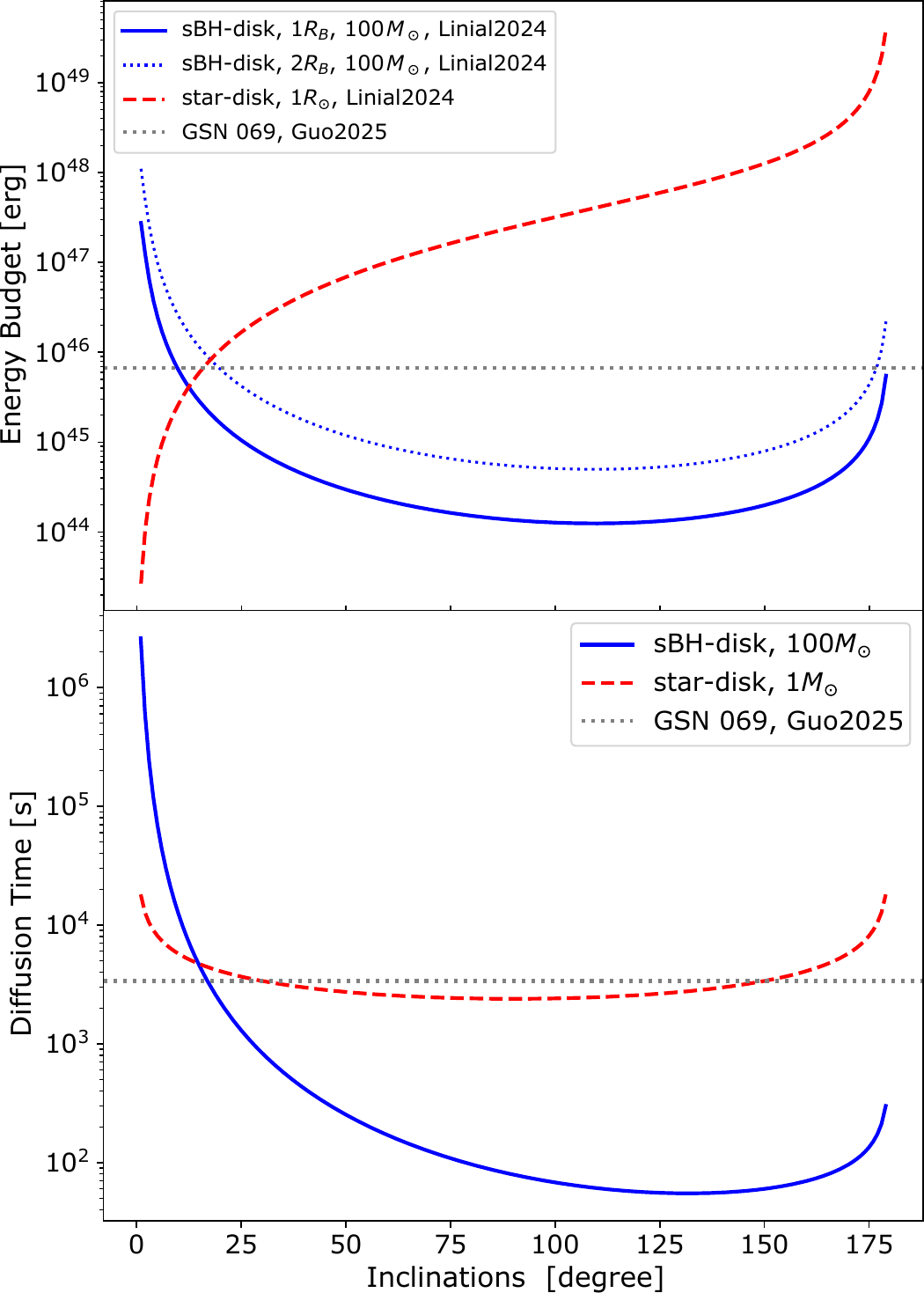}
\caption{Upper panel: Energy budget estimated with the Linial's model for a surface density of $10^6~\mathrm{g\,cm^{-2}}$. The blue solid and red dashed curves correspond to shocked regions of radius $R_{\mathrm{B}}$ and $2R_{\mathrm{B}}$. The gray dotted line indicates the minimum energy budget observed for GSN 069, $\simeq 7\times10^{45}\,\mathrm{erg}$ \citep{miniutti2023A&A...670A..93M, Guo2026ApJ...998...78G}. Lower panel: Duration time as a function of inclination. The blue curve shows the prediction of the sBH-disk model, while the red dashed curve corresponds to the star-disk model.
}
\label{fig:analysis}
\end{figure}

To estimate the energy budget in the sBH EMRI+disk model, we adopt a fiducial surface density of $10^6~\mathrm{g\,cm^{-2}}$ and a sBH of 100$M_{\odot}$ black hole. We assume a SMBH of $M_\bullet = 10^{6} M_\odot$ and an orbital period  $T_{\mathrm{orb}} = 63.7 \,\mathrm{ks}$, roughly consistent with GSN 069 \citep{miniutti2019Natur.573..381M}. For an inclination of $i=\pi/10$, the Bondi radius is $R_{\mathrm{B}} \approx 0.68 R_\odot$. For comparison, we also show the energy budget assuming an effective interaction radius of $2R_{\rm B}$, reflecting the possibility that the actual shocked region extends beyond the nominal Bondi radius (see Fig.~\ref{fig:analysis}, upper panel).

The energy input can also be reliably estimated from the gravitational drag force acting on the sBH, $F_{\mathrm{drag}}$, and the distance it travels through the disk. This approach is more accurate than the simple approximation $ M_{\mathrm{sh}} v_{\mathrm{rel}}^2$, because $M_{\mathrm{sh}}$ accounts only for the gas that could be accreted, representing just a fraction of the gas assumed to be shocked across the Bondi radius, whose actual extent typically exceeds the accretion radius by several times. Following \cite{zhou2024PhRvD}, the sBH energy loss can then be expressed as:

\begin{equation}
\begin{aligned}
E_{\mathrm{sBH}} &= F_{\mathrm{drag}} \cdot \Delta L \\
&= 4\pi \ln \Lambda \, \frac{G^2 m_{\mathrm{sBH}}^2}{v_{\mathrm{rel}}^2} \, \frac{\Sigma}{\cos(i/2)} \\
&\approx 2\times10^{46}\mathrm{ergs} \left(\frac{\Sigma}{10^{6}\mathrm{g\,cm^{-2}}}\right)\\
&\left(\frac{m_{\mathrm{sBH}}}{100\,M_{\odot}}\right)^{2}\left(\frac{\ln \Lambda}{5}\right)
\left(\frac{0.024}{\theta(i)}\right),
\end{aligned}
\end{equation}
where $\ln \Lambda$ is the Coulomb logarithm, which depends on the ratio of maximum to minimum impact parameters \citep{thun2016A&A...589A..10T}. Here $\theta(i)=\sin^{2}(i/2)\cos(i/2)$ encapsulates the inclination dependence of the relative velocity and the disk-crossing path length. At low inclinations, the reduced collision velocity enlarges the effective interaction radius, allowing the sBH to shock a substantial portion of the disk gas and produce QPEs. 
This mechanism also provides a plausible explanation for QPEs powered by sBH without invoking super-Eddington accretion \citep{lam2025PhRvD.112h3006L}, 
with the effective radius set by the Bondi radius but potentially extending up to the Hill radius. It should be noted that, in a thin-disk geometry, the disk thickness may be comparable to the Bondi radius or the size of the impacting object. Consequently, the Coulomb logarithm $\Lambda$ may approach or fall below unity, rendering the estimate unreliable. We therefore treat this expression as only a rough approximation and do not use it for quantitative predictions in this work. The formula is more appropriate for thick disks or situations closer to a uniform medium.

In addition to the energy budget, the burst duration of QPEs is also of interest and imposes an informative constraint on QPE models. 
The eruption duration is set when the photon diffusion timescale becomes comparable to the expansion timescale \citep{linial2023ApJ, vurm2025ApJ}，
\begin{equation}
\begin{aligned}
t_{\mathrm{e}} = \left(\frac{\kappa M_{\mathrm{sh}}}{4 \pi c v_{\mathrm{rel}}}\right)^{1 / 2} &\approx 4.3~{\mathrm{ks}} \left( \frac{\Sigma}{10^6~{\mathrm{g~cm^{-2}}}} \right)^{1/2} \left( \frac{R_\mathrm{eff}}{R_\odot} \right)  \\
&\left( \frac{v_{\mathrm{rel}}}{0.025c} \right)^{-1/2} \left( \frac{\kappa}{0.34~{\mathrm{cm^2~g^{-1}}}} \right)^{1/2}\ ,
\end{aligned}
\end{equation}
where $\kappa$ is the gas opacity.

Assuming the post-impact debris cloud expands approximately spherically, its optical depth rapidly decreases with time.
In the sBH-disk collision model, the duration time is highly sensitive to the orbital inclination (see Fig.~\ref{fig:analysis}, lower panel), which exhibits a dependence similar to that of energy budget. This strong sensitivity arises because the effective interaction radius is set by the Bondi radius, which depends on the relative velocity and hence on inclination, leading to the inclination scalings summarized below: 
\begin{equation}
t_{\mathrm{e,\,bh}} \propto \sin^{-\frac{5}{2}}(i/2)\,\cos^{-\frac{1}{2}}(i/2).
\end{equation}
For a surface density of $10^6 \, \mathrm{g \, cm^{-2}}$, the duration at an inclination of $\pi/10$ is about 3 ks, which is close to the observed properties of GSN 069. By contrast, in the star-disk model the effective interaction radius is fixed by the stellar radius and is independent of the relative velocity, leading to a much weaker inclination dependence, and hence durations confined to a narrower range of $\sim $3--13 ks. This inclination scaling can be written as:
\begin{equation}
t_{\mathrm{e,\,st}} \propto \sin^{-1/2}(i). \label{eq:t_e,st}
\end{equation}
The strong inclination sensitivity of the sBH-disk case, spanning nearly three orders of magnitude at low inclinations (Fig. \ref{fig:analysis}), may therefore provide a natural explanation for the large diversity of observed QPE durations. While inclination does play a role, other physical factors also contribute to this diversity. These include variations in local gas density, disk scale height, opacity, and ejecta geometry, as well as additional mechanisms such as disk precession, warping, and delayed emission following the impact.

It is worth noting that observations reveal an approximately linear scaling between flare duration and recurrence time \citep{arcodia2025ApJ...989...13A, baldini2026A&A...706L..15B}. Satisfying this trend does not require systematically lower inclinations at longer periods, as would be expected from the stellar model (Eq. \ref{eq:t_e,st}). This will be examined in detail using the semi-analytical model below (Section \ref{subsec:sBH-disk}).

In the star-disk model, the burst energy arises primarily from shock compression of the disk gas upon impact, whereas in the sBH-disk interaction, gravitational effects dominate the energy transfer. Numerical simulations have demonstrated that as a black hole traverses the disk, it perturbs the surrounding medium through gravitational drag within its Bondi-Hoyle radius \citep{ivanov1998ApJ...507..131I}. The black hole thereby injects part of its orbital kinetic and gravitational potential energy into the disk material, leading to localized heating and substantial radiation output, with our derivation indicating that this mechanism is velocity- and inclination-sensitive. Given that the Eddington luminosity of a stellar-mass black hole is far below the typical luminosities observed in QPEs, accretion-powered emission can be safely neglected in this scenario.

In the following section, we employ global hydrodynamic simulations to investigate both the star-disk and sBH-disk collision models, allowing us to compare their model predictions.

\section{Simulation Setup} \label{sec:sim}

In this study, we perform global three-dimensional hydrodynamical simulations of star-disk and sBH-disk collisions using the meshless finite-mass (MFM) method implemented in the \texttt{GIZMO} code \citep{hopkins2015MNRAS.450...53H}.  By computing inter-particle fluxes via Riemann solvers, it can accurately resolve shocks generated during collisions, while also delivering higher computational efficiency compared to alternative SPH methods \citep{hopkins2015MNRAS.450...53H}. The MFM method maintains second-order consistency in smooth flows, ensuring fast and reliable convergence for both shock capturing and smooth-field evolution \citep{deng2019local}. The particle-based method readily allows for global simulations that naturally treat all the gravitational interactions among the SMBH, disk, and star, including their self-gravity \citep{zhang2025tidal}.

\begin{figure*}[ht!]
\centering
\includegraphics[width=0.67\textwidth]{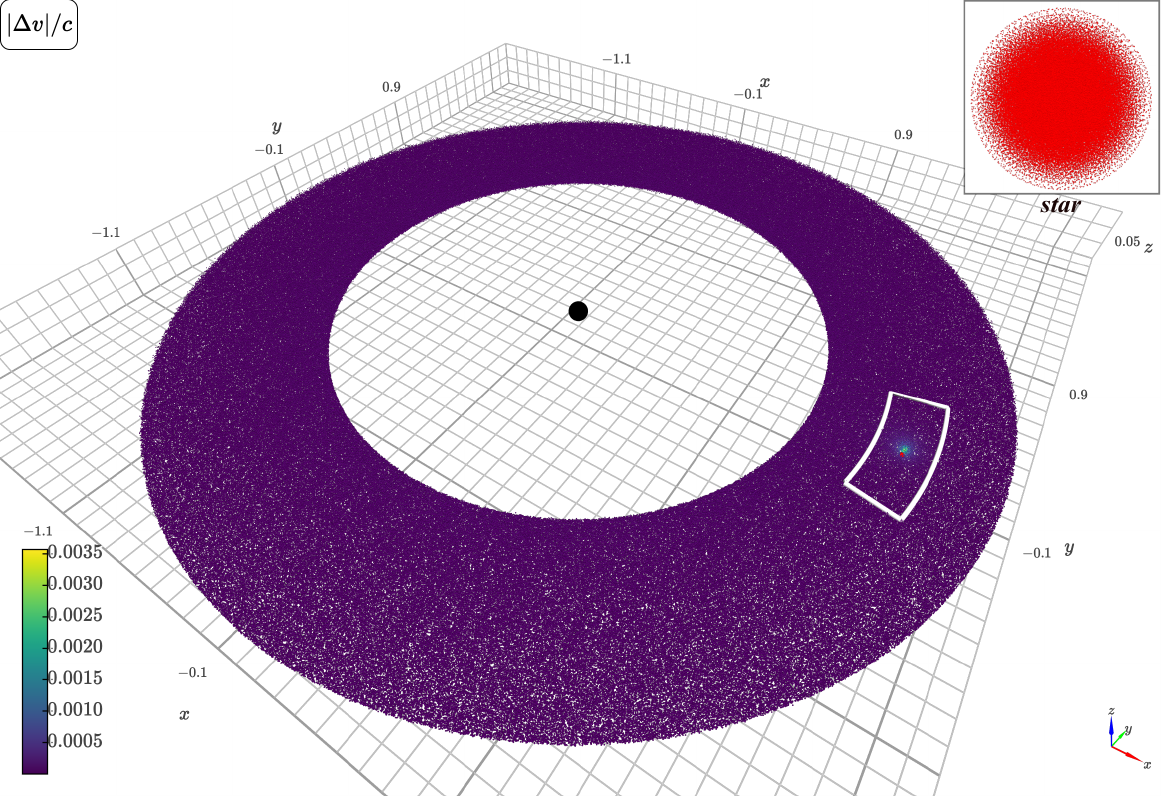}
\caption{Global simulation of a full accretion disk impacted by an EMRI. The color scale shows the absolute velocity perturbation $  |\Delta v|/c $ relative to the unperturbed disk. Regions outside of the Hill sphere remain essentially unaffected by the intruder. The white box outlines the simulation domain we adopted in our production runs. This sector has a radial width of $\sim 0.2~\mathrm{AU}$, and spans an azimuthal angle corresponding to $\sim 1/16$ of the local disk circumference, centered on the impact point. The upper-right panel displays a magnified view of the stellar EMRI resolved by $\sim10^5$ MFM particles in our simulations.}
\label{fig:global}
\end{figure*}

We assume a polytropic equation of state with $\gamma=4/3$ for the gas disk around an SMBH, appropriate for radiation-pressure-dominated gas. Radiative transfer is not included in current simulations. As shown in Fig. \ref{fig:global}, the impact affects only a limited portion of the disk that is strictly confined within the Hill radius of the secondary. We can therefore safely restrict the computational domain to a sufficiently large annular sector (the region enclosed by the white solid line in Fig. \ref{fig:global}), whose radial and azimuthal extent fully covers a circular area comparable in size to the secondary's Hill radius. The reduced domain allows us to maintain high resolution around the interaction while preserving the full gravitational potential of the central SMBH, which is essential for accurately capturing the gas dynamics across the Bondi-to-Hill scales. The vertical scale height is fixed to $1\,R_\odot$ for both numerical convenience and structural stability, corresponding to a relatively compact disk geometry. The portion of the disk is resolved by $\sim5\times10^6$ particles, i.e., at a mass resolution of $\sim 1.6\times10^{-9} M_{\odot}$ per particle. Such a mass resolution is sufficient to characterize the ejecta (which typically contain $\sim10^4$ particles), allowing us to reliably measure both the total ejecta mass and its morphological structure. To further verify numerical convergence, we carried out additional simulations at lower resolutions of $\sim 1 \times 10^6$ and $\sim 3 \times 10^6$ particles. The total ejecta mass in these lower-resolution runs differs from the fiducial run by approximately 9.7\% and 1.3\%, respectively. The substantial reduction in the discrepancy from $1\times10^6$ to $3\times10^6$ particles demonstrates clear convergence, indicating that further increases in resolution are unlikely to significantly alter the main results.

We also conduct stability tests to ensure that, compared with the initially prescribed disk configuration, no perturbations large enough to affect the subsequent collision simulations are produced. In the relaxation run without collisions, the disk retains a sharp edge and stable structure after 2 hours---twice the duration of the collision simulations-demonstrating its stability. 

For the stellar case, we model a self-gravitating polytropic star with solar mass and radius following the Lane-Emden solution with an index corresponding to $\gamma = 5/3$, represented by $\sim10^{5}$ particles (see the upper right corner of Fig.~\ref{fig:global}). This resolution is sufficient to capture the essential interactions with the surrounding disk particles, given that our primary focus is on the gas dynamics and the propagation of perturbations rather than the detailed stellar interior. The stellar-mass black hole, in contrast, is treated as a point mass, since its accretion contributes negligibly to the overall energy budget, the accretion process is neglected.

We focus on the evolution of the accretion disk following the impact by two types of secondary objects-a $100\,M_\odot$ black hole and a $1\,M_\odot$ star-traversing the disk with orbital parameters that are broadly consistent with GSN 069. Since radiative cooling is not included in our simulations, the post-shock gas retains its thermal energy for an extended period, potentially leading to more pronounced expansion in the later stages than would occur in a realistic cooling environment. The simulations cover a range of disk surface densities ($10^4$--$10^6\,\mathrm{gcm^{-2}}$) and different inclination angles ($\pi/10$ and $\pi/2$). Each collision is followed for approximately one hour of simulated time (typical QPE duration of GSN 069), capturing the disk's immediate reaction to the perturbers, including the propagation of perturbations and the resulting changes in flow structure and density distribution.

\section{Results} \label{sec:results}

\begin{deluxetable*}{ccccccc}
%\digitalasset
\tablewidth{0pt}
\tablecaption{Simulation Results \label{tab:table}}
\tablehead{
\colhead{case} & \colhead{Inclination} & \colhead{$\Sigma\,(\mathrm{g\,cm^{-2}})$} &\colhead{$M_\mathrm{sh}^\mathrm{f}$/$M_\mathrm{sh}^\mathrm{b}$ ($M_\odot$/$M_\odot$)}  & \colhead{$v_\mathrm{ej}^\mathrm{f}$/$v_\mathrm{ej}^\mathrm{b}$ (c/c)} & \colhead{$E_\mathrm{sh}^\mathrm{f}$/$E_\mathrm{sh}^\mathrm{b}$ (erg/erg)} & $R_{\mathrm{B}}(\mathrm{or}\, R_{\mathrm{*}})/R_{\mathrm{H}}$
}
\startdata
$bh\_i18\_sigma1e6$ & $\pi/10$ & $10^{6}$ & $3.2\times10^{-5}/2.9\times 10^{-5}$ & 0.029/0.022 & $2.38\times 10^{46}/1.21\times 10^{46}$ & 0.061\\
$bh\_i90\_sigma1e6$ & $\pi/2$ & $10^{6}$ & $5.6\times 10^{-6}/5.2\times 10^{-6}$ & 0.008/0.008 & $3.21\times 10^{44}/2.98\times 10^{44}$ & 0.003\\
$bh\_i18\_sigma1e5$ & $\pi/10$ & $10^{5}$ & $3.3\times 10^{-6}/3.4\times 10^{-6}$ & 0.029/0.022 & $2.45\times 10^{45}/1.41\times 10^{45}$ & 0.061\\
$bh\_i18\_sigma1e4$ & $\pi/10$ & $10^{4}$ & $3.2\times 10^{-7}/2.8\times 10^{-7}$ & 0.029/0.022 & $2.38\times 10^{44}/1.16\times 10^{44}$ & 0.061\\
$st\_i18\_sigma1e6$ & $\pi/10$ & $10^{6}$ & $2.0\times 10^{-5}/3.2\times 10^{-6}$ & 0.024/0.010 & $1.10\times 10^{46}/2.74\times 10^{44}$ & 0.421\\
\enddata
\tablecomments{ Here ``$st/bh$'' indicates the impactor type, ``$i$'' the orbital inclination, and ``$sigma$'' the disk surface density. The ejecta mass is evaluated one hour after the impact as the mass of material located above a height of 4H from the disk mid-plane, and the corresponding energy budget is estimated as $\frac{1}{2} M_\mathrm{sh} v_\mathrm{ej}^2$, where $v_\mathrm{ej}$ is the bubble expansion velocity measured in the simulation.}
\end{deluxetable*}

In this section, we examine how the passage of secondary objects-either a solar-mass star or a stellar-mass black hole-through an accretion disk affects the disk gas. Hydrodynamic simulations provide a controlled setting to systematically explore the influence of different orbital parameters and disk properties on the resulting gas dynamics. Particular attention is given to the morphology, mass, and velocity distribution of the ejecta, as well as the formation of bow shocks and the resulting flow structures. We estimate key parameters (see Table \ref{tab:table}), such as the energy budget, to evaluate whether these processes can reproduce the observed features of QPEs. 

\subsection{Star-Disk collision}

\begin{figure*}[ht!]
\centering
\includegraphics[width=0.8\textwidth]{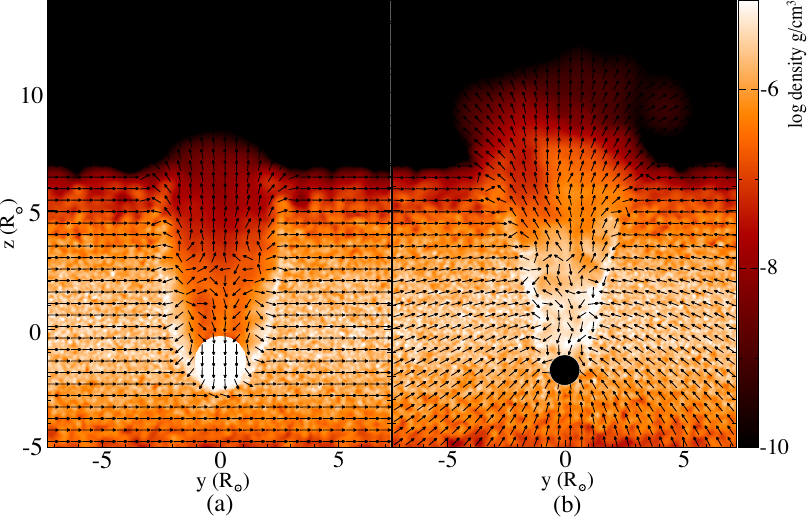}
\caption{Vertical slices through the disk along the azimuthal direction showing the density distribution and velocity field. Arrow lengths are not to scale. (a) The bow shock is generated by a star-disk collision, and downstream material fills the low-density cavity left by the star's passage, producing backward ejecta; (b) The perturbation induced by the sBH-disk interaction, exhibiting a bow-shock-like structure. The black hole attracts surrounding material along its path, producing a pronounced density enhancement and quickly generates backward ejecta. (The size of black-filled part is bondi radius of sBH.)}
\label{fig:bowshock}
\end{figure*}

From the simulations, we measured the ejecta mass and roughly estimated the energy, which are listed in Table \ref{tab:table} for comparison with theoretical predictions. Here we focus on sBH-disk collision while we performed one star-disk collision simulation ($st\_i18\_sigma1e6$) to highlight its difference to sBH-disk collisions. In Fig.~\ref{fig:bowshock}(a), we visualize the flow density near the impact point, with arrows indicating the flow velocity. During the impact, a bow shock forms ahead of the star and propagates horizontally, perturbing gas within twice the stellar radius significantly. We define the forward ejecta as moving roughly along the impactor's trajectory, while the backward ejecta moves oppositely. At the star passage, part of the gas is directly ejected out of the disk, while a low-density wake forms along the star's trajectory. This region is rapidly refilled with shock-heated gas driven in by strong pressure gradients from both sides, generating backward-directed ejecta.

\begin{figure*}[ht!]
\plotone{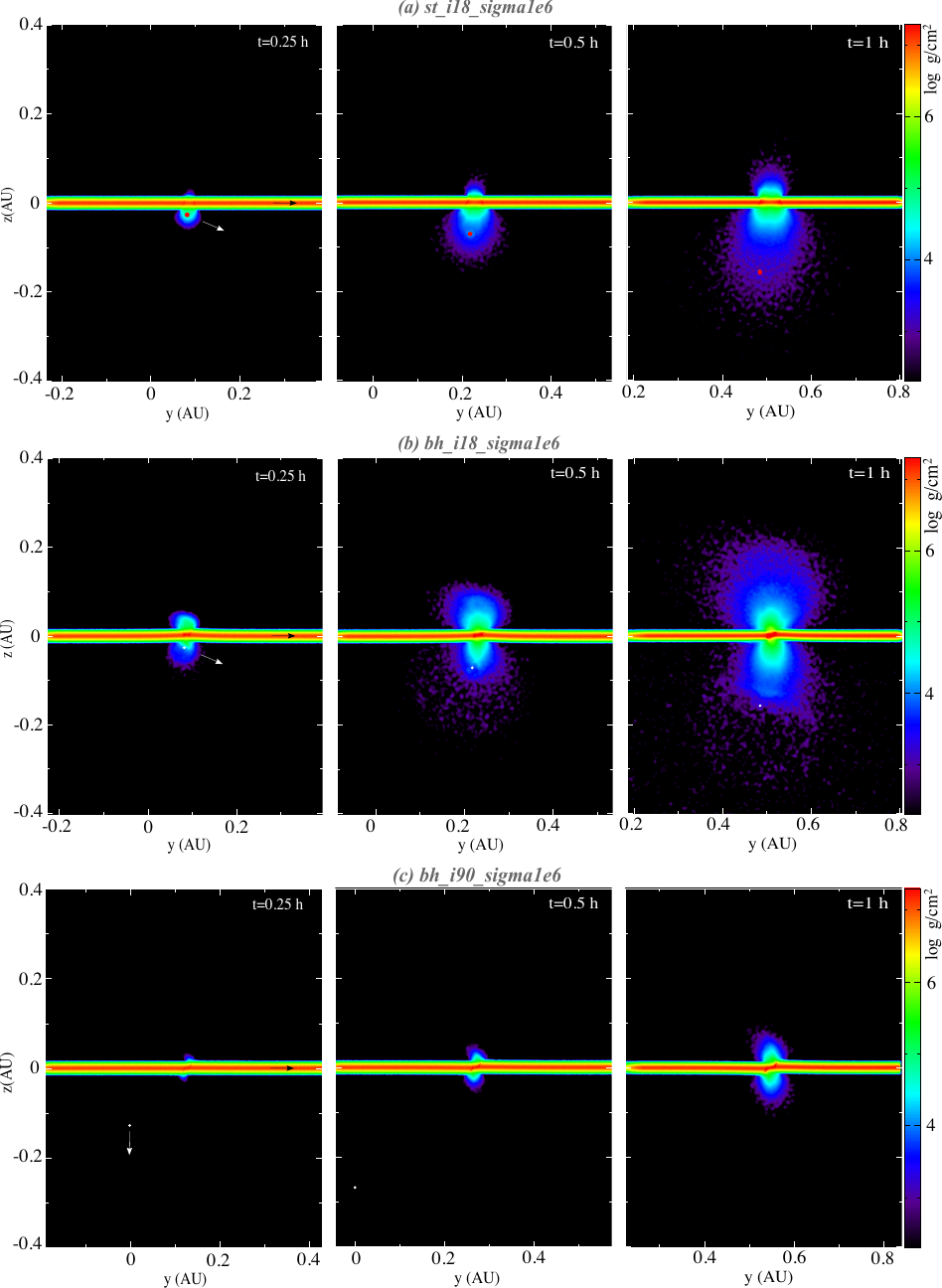}
\caption{Projected density plot for three simulations in Table \ref{tab:table}, which involve stellar or black hole impacts on the disk at inclinations of $\pi/10$ and $\pi/2$. In the first column, arrows mark the directions of motion of both the disk gas and the impactor. For each simulation, the snapshots are shown at 0.25, 0.5, and 1.0 hours after the initial contact. 
}
\label{fig:sim123}
\end{figure*}

However, the mass of this backward ejecta is far less than the forward component ($M_\mathrm{sh}^\mathrm{f}/M_\mathrm{sh}^\mathrm{b} \gtrsim 6.3$ in $st\_i18\_sigma1e6$), indicating that the forward outflow dominates both the mass and energy output. This strong asymmetry mainly originates from the physical size of the star, which blocks the downstream flow and produces a rarefied wake region with a density significantly lower than the unperturbed disk. Moreover, the bulk of the forward ejecta moves at nearly the relative velocity $  v_\mathrm{rel}  \simeq 0.025 \,c$, while the bulk of the backward ejecta expands at less than half that speed (see Table \ref{tab:table} and Figure \ref{fig:sim123} (a)). This pronounced velocity asymmetry highlights that the simplifying assumption commonly adopted in analytic models, that ejecta on both sides move at approximately the relative velocity, is idealized and may not hold for stellar impacts. In terms of time evolution, the forward ejecta diffuse rapidly and produce sharply peaked energy-release curves, while the backward component contributes more smoothly and over a longer timescale. Consequently, the luminosity becomes increasingly dominated by the forward ejecta, highlighting a forward-backward asymmetry that poses a `symmetry problem' for the star-disk collision model as a viable explanation of QPEs. 

Tests with other inclination angles confirm that the strong asymmetry in the ejecta persists. This behavior is consistent with the findings of \cite{huang2025ApJ...993..186H}. The authors argue that the strong asymmetry between forward and backward ejecta is most pronounced in the bolometric light curves and in lower-energy bands such as UV and optical, while this asymmetry may be reduced in the soft X-ray band in which QPEs are detected, because the optical depth at soft X-ray frequencies is much lower than the frequency-integrated optical depth used for bolometric calculations.

A crucial difference from previous local simulations \citep{yao2025ApJ...978...91Y, huang2025ApJ...993..186H} is that our global setup includes the full gravitational potential of the central SMBH. The Hill sphere of the stellar EMRI is restricted to only $\sim$ 2--3 times larger than the stellar radius, which is comparable in size to the bow-shock region. As a result, gravitational perturbations are highly localized, and the disk material remains essentially undisturbed Keplerian motion outside the bow-shock region (one can compare our Fig. \ref{fig:bowshock}(a) with Fig. 10 in \citet{yao2025ApJ...978...91Y} and Fig. 2 in \citet{huang2025ApJ...993..186H}). 

Finally, our primary focus is the response of the disk gas to the impact under realistic global conditions. The detailed post-impact evolution of the stellar structure itself is not the main subject of this work, as the stellar resolution ($  \sim 10^5  $ particles) is insufficient to capture subtle internal changes; this will be explored in future studies.

\begin{figure*}[ht!]
\plotone{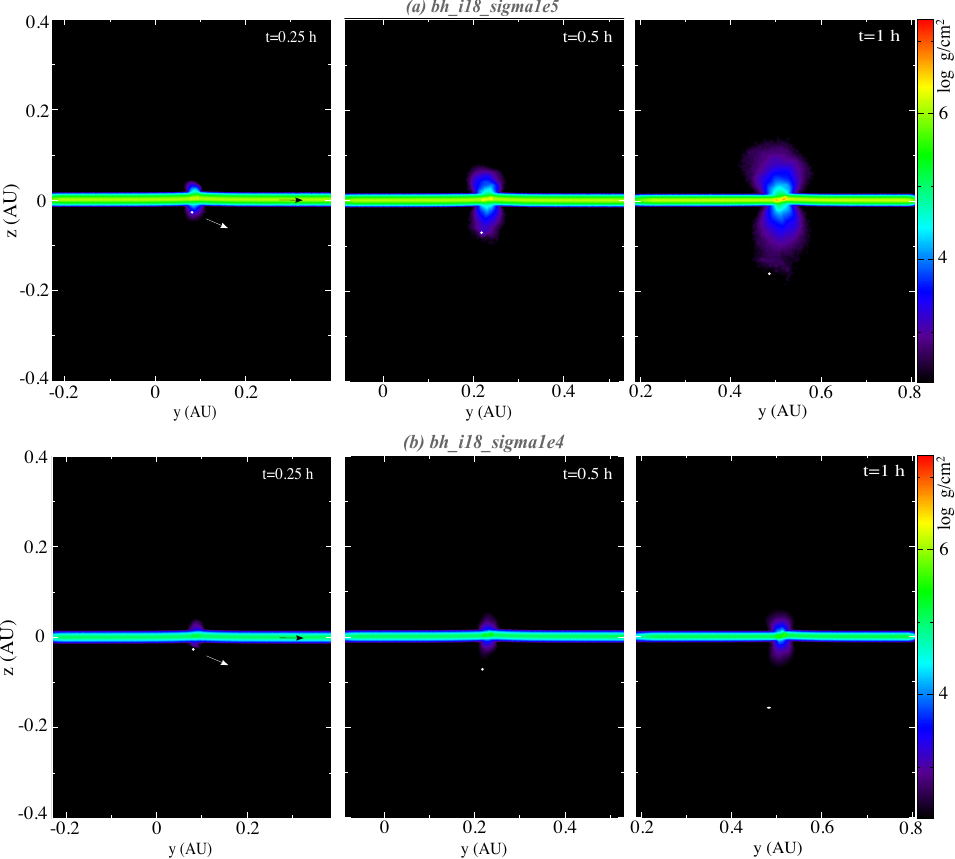}
\caption{Density projection snapshots for simulations with reduced disk surface densities, compared with the reference case $bh\_i18\_sigma1e6$. These runs investigate black hole impacts on accretion disks with lower surface densities of $10^{5}$ and $10^4\,\mathrm{g\,cm^{-2}}$, while all other parameters are kept identical to those in the reference simulation.
}
\label{fig:sim45}
\end{figure*}

\subsection{sBH-Disk collision}

In Fig. \ref{fig:bowshock}, we show that the interaction of a $100\,M_\odot$ black hole with the accretion disk differs from the star-disk case in two fundamental ways. First, the black hole has a negligible physical cross-section and therefore does not obstruct the downstream flow, allowing the wake region to retain a density comparable to or higher than the unperturbed disk. Second, a $100\,M_\odot$ black hole possesses a much larger Hill radius ($  \sim 10\,R_\odot  $) than a solar-mass star, enabling it to gravitationally capture and perturb a much greater volume of disk gas. Together, these effects produce significantly larger and far more symmetric ejecta masses on both sides.

This improved symmetry is clearly visible in the low-inclination run $  bh\_i18\_sigma1e6  $ (Fig. \ref{fig:sim123} (b)): the bulk of the ejecta on both sides moves at velocities comparable to the relative velocity ($  v_\mathrm{rel} \approx 0.025\,c  $), producing nearly symmetric outflows. A mild asymmetry arises only from a fast-expanding outer envelope that reaches up to $  \sim2.4\,v_\mathrm{rel}  $ forward and $  \sim1.7\,v_\mathrm{rel}  $ backward. However, this component comprises only $  \sim10\%  $ of the total ejecta mass, so the bulk of the two-sided ejecta remains nearly symmetric. This stands in sharp contrast to the stellar run $  st\_i18\_sigma1e6  $ (Fig. \ref{fig:sim123} (a)), where the forward ejecta expands nearly twice as fast as the backward component.

The ejecta velocity behavior in the sBH-disk case arises from a fundamentally different mechanism than in the star-disk case. In star-disk collisions, the ejecta is primarily driven by \textit{shock compression} as the star plows through the disk; therefore, the bulk ejecta velocity naturally tracks the relative velocity $  v_{\rm rel}  $. In contrast, in sBH-disk collisions the ejecta is driven mainly by \textit{gravitational pull} exerted by the black hole on the surrounding disk gas within its effective interaction region (jointly set by the Bondi and Hill radii).

At low inclination ($  i=\pi/10  $), the Bondi radius is relatively large ($  \sim 0.68\,R_\odot \approx 0.06\,R_\mathrm{H}  $). This allows a greater mass of disk gas to be gravitationally focused and accelerated over a longer interaction time, resulting in a higher ejecta velocity. At high inclination ($  i=\pi/2  $), the Bondi radius shrinks dramatically ($  \sim0.03\,R_\odot \approx 0.003\,R_\mathrm{H}  $), so significantly less gas experiences strong gravitational acceleration. Consequently, the bulk ejecta velocity drops to a much lower value ($  \sim 0.008\,c  $) while remaining nearly symmetric on both sides. In this regime, the ejecta velocity is no longer comparable to the (much higher) relative velocity (also see the discussion in Section \ref{subsec:sBH-disk}).

Our results also show that the ejecta mass scales linearly with the disk surface density. Comparing the three low-inclination runs ($  bh\_i18\_sigma1e6  $, $  bh\_i18\_sigma1e5  $, and $  bh\_i18\_sigma1e4  $), we find that both the forward and backward ejecta masses decrease proportionally with $  \Sigma  $. However, the measured ejecta mass exceeds the analytic predictions based only on the Bondi radius (Section \ref{sec:model}). For example, in the fiducial case with $  \Sigma = 10^6\,\mathrm{g\,cm^{-2}}  $, the ejecta mass on either side reaches $  \sim 3 \times 10^{-5}\,M_\odot  $, which significantly exceeds the analytic estimate of $  \sim 3.6 \times 10^{-6}\,M_\odot  $ obtained by assuming the effective radius equals the Bondi radius alone. We further infer the effective interaction radii are $  \sim2  $--$3\,R_{\rm B}$ ($  \approx 0.18  $--$0.27\,R_{\rm H}$) at low inclination ($  i=\pi/10  $) and  $  \sim20\,R_{\rm B} $ ($\approx 0.06\,R_{\rm H}  $ ) at high inclination ($  i=\pi/2  $).

To summarize, these results demonstrate that a stellar-mass black hole can generate sufficient energy and produce sufficiently symmetric ejecta to power observed QPEs, particularly at low orbital inclinations. The key difference from previous analytic estimates lies in the effective interaction scale, which extends well beyond the Bondi radius and is naturally bounded by the Hill radius in the presence of the central SMBH’s tidal field. This gravitational-drag mechanism, combined with the strong inclination dependence, allows the sBH-disk model to accommodate a wide range of QPE energies and durations without requiring intermediate-mass black holes. These findings motivate a more quantitative semi-analytical treatment, which we develop in Sections \ref{subsec:Reff} and \ref{subsec:sBH-disk}.

\section{Discussion} \label{sec:discussion}
In this section, we discuss the applicability and limitations of the star-disk and sBH-disk models in light of our simulation results and observational constraints, some of which have already been mentioned in Sections \ref{sec:results}. We also outline potential directions for future observational and theoretical investigations.

\subsection{Challenges for the Star-Disk Collision Model}

Our simulations confirm that star-disk collisions tend to produce strongly asymmetric eruptions. However, the observed strong -- weak pattern does not involve such a large contrast; in RX J1301.9+2747, for example, the stronger flare exceeds the weaker one by only a factor of $\sim2$--$3$. This behavior can be explained more naturally in a sBH-disk system, where orbital eccentricity and impact orientation play key roles. In particular, the highly eccentric orbit of RX J1301.9+2747 likely results in varying impact angles and local disk conditions between successive encounters, producing more noticeable contrasts than in other QPEs.

\begin{figure}[ht!]
\centering
\includegraphics[width=\columnwidth]{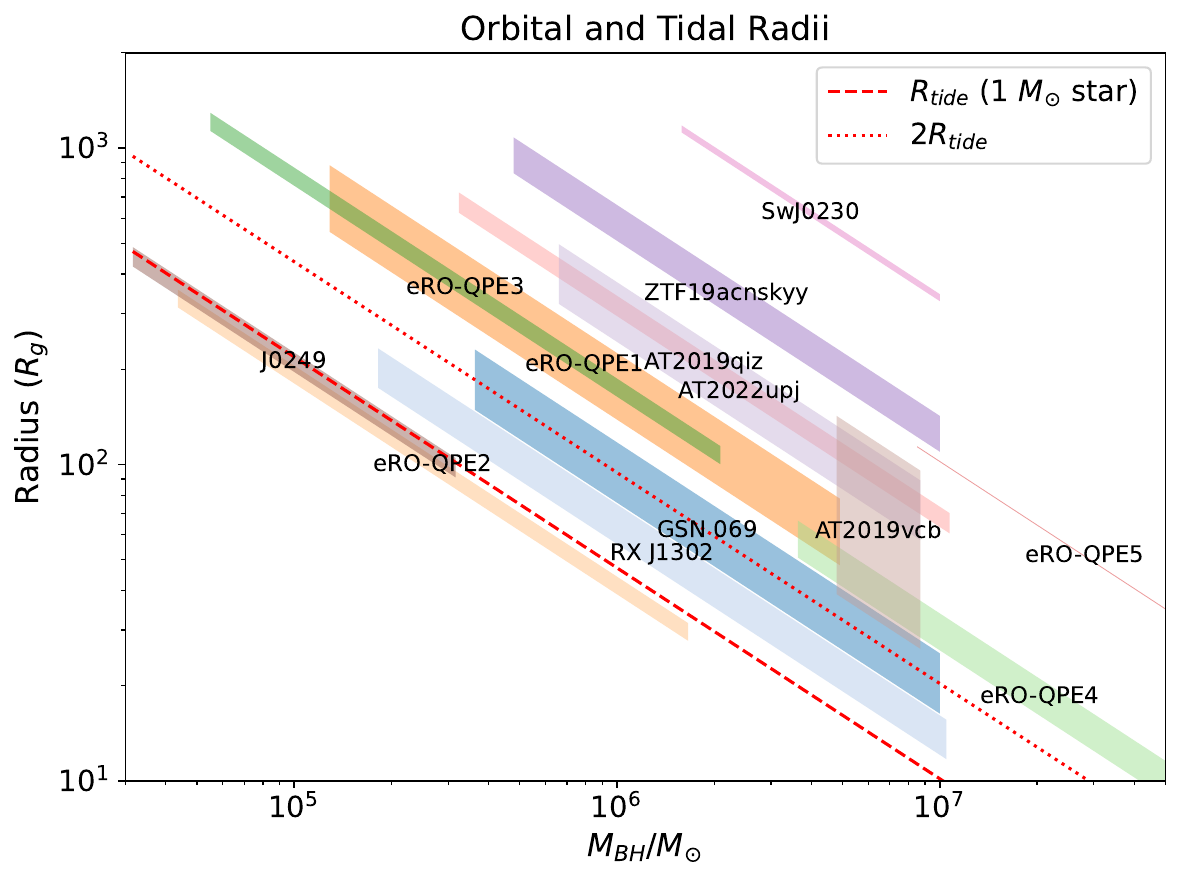}
\caption{Figure showing the relationship between the circular orbital radius and the tidal radius for a $1\,M_\odot$ star as the impactor. According to the simulation results, in most cases only the brighter of the two eruptions is observable, so the observed QPE period effectively corresponds to the star's orbital period ($T_{\mathrm{orb}}=T_{\mathrm{QPE}}$).}
\label{fig:tidalradii}
\end{figure}

Our simulations suggest that the backward ejecta produce eruptions that are fainter by more than an order of magnitude than the forward component and are therefore likely unobservable. Consequently, if only the forward flare were detected, implying that the observed QPE recurrence time is effectively equal to the stellar orbital period. Under this assumption, additional physical mechanisms are required to explain the commonly observed alternating strong-weak amplitude pattern and the characteristic long-short recurrence intervals.

Adopting the observed QPE recurrence time as the stellar orbital period introduces yet another major difficulty for the star-disk model. A solar-mass main-sequence star is expected to be fully disrupted if its orbit lies within roughly twice the tidal disruption radius \citep[see e.g., ][]{2011ApJ...732...74G, 2013ApJ...762...37L}. Comparing the inferred orbital semi-major axes of known QPE sources with this limit (Fig.~\ref{fig:tidalradii}) reveals that most are safely outside the danger zone, but eRO-QPE2, XMMSL J024916.6–041244, and RX J1301.9+2747 all lie within $2R_{\rm tide}$, while GSN 069 sits very close to the threshold. Even under the conservative assumption of a circular orbit with a period twice the observed QPE recurrence time (i.e., neglecting the unobservable backward eruption), two short-period sources remain within twice the tidal radius. This tidal stability problem is therefore severe for a solar-mass star and can only be partially alleviated by adopting a lower-mass impactor.

Radiation-transport simulations by \citep{vurm2025ApJ} further highlight the tension. They find that reproducing the observed soft X-ray luminosities and temperatures in the photon-starved regime requires either high collision velocities ($  v_{\rm coll} \gtrsim 0.15c  $) or significantly inflated stellar radii ($  R_\star \gtrsim 10{-}50\,R_\odot  $), possibly resulting from repeated collisions. However, such large or inflated stars would be even more vulnerable to tidal disruption for QPE sources at the small orbital separations.

The stellar debris stream model \citep{yao2025ApJ...978...91Y,linial2025ApJ...991..147L} suggests that repeated stellar impacts can drive significant mass loss and expansion of the stellar envelope, forming elongated debris streams that subsequently collide with the disk and may power the observed QPEs (including the the backward ejecta). While this is an interesting possibility, a major challenge for any stellar (or debris-stream) scenario is the limited lifetime of the star under repeated impacts, which requires further verification of  its feasibility. Additionally, \citet{yao2025ApJ...978...91Y, huang2025ApJ...993..186H, vurm2025ApJ} suggest that a thicker disk may produce more symmetric ejecta. We also tested a disk with a scale height three times larger than our fiducial model and found that the asymmetry of the ejecta is not mitigated, although the time delay between the ejecta emerging from the two sides becomes larger. Recent work by \citep{2026arXiv260400953H} suggests that oblique star-disk collisions can significantly reduce the forward-backward ejecta asymmetry compared to perpendicular impacts, potentially alleviating one of the main difficulties for the star-disk model.

\subsection{Revisiting the Effective Interaction Radius}
\label{subsec:Reff}

\begin{figure*}[ht!]
\plotone{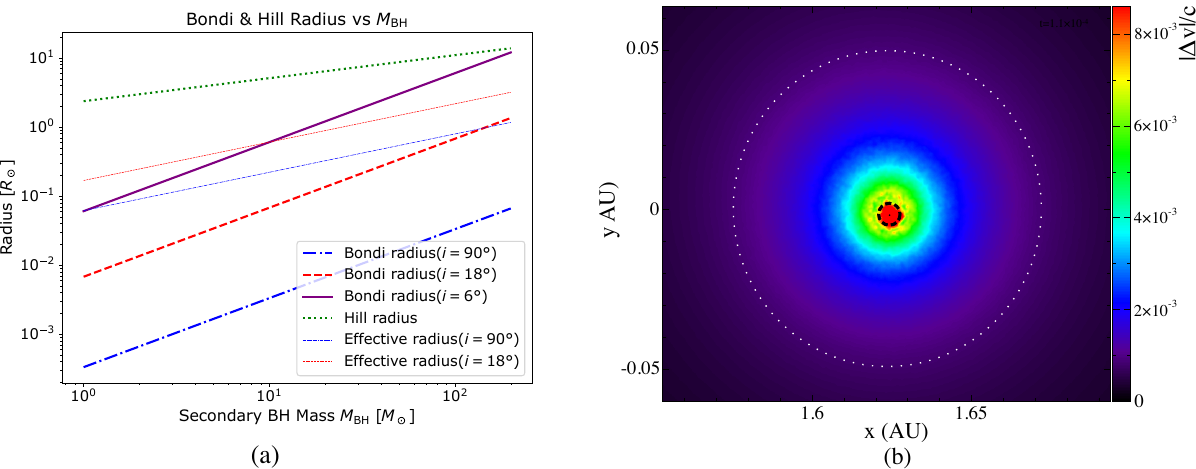}
\caption{(a) Comparison of Bondi and Hill radii under different impactor masses and orbital inclinations, for a central black hole mass $M_\mathrm{BH}=1.05\times10^6\,M_\odot$ and a semi-major axis $a=1.62\,\mathrm{AU}\, (\approx 160\,R_{g})$. Effective radii $  i=\pi/10  $ and $  i=\pi/2  $ are shown for reference; (b) Face-on view of the $|\Delta v|$ distribution from the $bh\_i18\_sigma1e6$ simulation at the moment when the black hole passes through the disk mid-plane. The white dotted circle denotes the Hill radius, and the black dashed circle shows the Bondi radius for reference.}
\label{fig:eff_radius}
\end{figure*}

Our global 3D hydrodynamic simulations demonstrate that simply adopting the Bondi radius as the effective interaction radius $R_\text{eff} \sim R_{\rm B}$, a common assumption in previous analytic work (see Section \ref{sec:model}), severely underestimates the true scale of the interaction between a stellar-mass black hole ($\sim 100\,M_\odot$) and the accretion disk.

To determine how the effective interaction radius relates to the Bondi and Hill radii, in Fig.~\ref{fig:eff_radius}(a) we first show how the Bondi radius $R_\mathrm{B}$ depends on black-hole mass and orbital inclination (since inclination directly controls the relative velocity between the secondary and the Keplerian disk gas), while the Hill radius $R_\mathrm{H}$ depends only on mass and orbital separation (both are fixed in our simulations in this work). One can notice that for a $100\,M_\odot$ black hole, the Hill radius is always larger than the Bondi radius under the physical conditions considered here. And different inclinations can result in variations of Bondi radius by almost two orders of magnitude. Here we mainly discuss low ($  i=\pi/10  $) and high ($  i=\pi/2  $) inclinations, and extreme low inclination ($  i=\pi/30  $) is only for illustration purposes.

In Fig. \ref{fig:eff_radius}(b), we further illustrate the relative strength of the gravitational pull exerted by the secondary across different radial zones (run $bh\_i18\_sigma1e6$). The pull is strongest within and near the Bondi radius (black dashed circle) and decays rapidly outward, yet remains dynamically significant throughout the entire Hill sphere (white dotted circle). This qualitative pattern holds at all inclinations, but at lower inclinations the Bondi radius is larger, so the region of strong gravitational influence extends over a correspondingly wider radial range.

In order to quantify the size of the shocked region in a simple and practical way, we propose the following empirical scaling of the effective radius $  R_{\rm eff}  $ as a replacement for the Bondi radius, which offers a better approximation to the interaction scale than using $R_\mathrm{B}$ alone: 
\begin{equation}\label{eq:R_eff}
R_{\mathrm{eff}} \simeq 0.5 \,R_\mathrm{B}^{1/3} \, R_{\mathrm{H}}^{2/3}.
\end{equation}
This ad hoc relation is not rigorously derived from the simulations but serves as a guess that balances contribution from the Bondi and Hill radii, reproducing the measured ejecta mass and energy budgets more accurately than the estimation with the Bondi radius (see Subsection \ref{subsec:sBH-disk} for details). 
The effective interaction radius $R_{\rm eff}$ discussed here represents the scale over which the secondary can significantly perturb, heat, and entrain disk gas during a crossing, leading to observable ejecta. It is physically distinct from the classical Bondi-Hoyle-Lyttleton (BHL) accretion radius. Our simulations show that this scale extends beyond the simple Bondi radius and is naturally bounded by the Hill radius in the presence of the central SMBH’s tidal field. The empirical relation in Eq.~(\ref{eq:R_eff}) is therefore a fit valid within the tidal regime explored in this work. We caution that it should not be extrapolated to the idealized no-tide limit, where the Hill radius loses its physical meaning as a cutoff. This is also reflected in local simulations \citep[e.g.,][]{lam2025PhRvD.112h3006L}, where the interaction scale is truncated by an arbitrarily chosen computational domain rather than by a physical boundary, unlike the Hill radius that naturally arises in global tidal setups.

This empirical relation shows that the effective radius is primarily controlled by the Hill radius (exponent 2/3), which was completely ignored in previous analytic work. As a result, earlier theoretical calculations severely underestimated the shocked mass produced by an sBH, as they relied solely on the Bondi radius. However, for a given secondary mass and orbital separation, the Hill radius is essentially fixed, so most of the variation in $  R_{\rm eff}  $ actually comes from changes in the Bondi radius. Although the exponent in $  R_{\rm B}  $ is only 1/3, the Bondi radius itself spans more than two orders of magnitude across the relevant range of inclinations (see Fig. \ref{fig:eff_radius} (a)). This large dynamic range means that even the weaker dependence on $  R_{\rm B}  $ produces a major impact on the magnitude of the effective interaction radius, therefore on the mass and flare energy of the shocked gas.

In Fig. \ref{fig:eff_radius}(a), we show the effective radius as a function of sBH mass for inclination at $  i=\pi/10  $ and $  i=\pi/2  $, respectively. For example, at low inclination ($  i=\pi/10  $), the effective radius is $  R_{\rm eff}\approx 3.2\,R_{\rm B} \approx 0.20\,R_{\rm H} \approx 2.18\,R_\odot  $. At high inclination ($  i=\pi/2  $), the ratio $  R_{\rm eff}/R_{\rm B}  $ becomes dramatically larger ($  \approx 24  $), but the absolute effective radius is actually smaller $  R_{\rm eff}\approx 24\,R_{\rm B} \approx 0.07\,R_{\rm H} \approx 0.8\,R_\odot  $. 

\subsection{Semi-Analytical Model of sBH-Disk Collisions}
\label{subsec:sBH-disk}

In light of results of our global simulations, we develop a semi-analytical model for the shocked mass and gravitational energy gain during the sBH traversal. The model assumes a thin disk with constant scale height $  H = 1 R_\odot  $ and surface density $  \Sigma = 10^6  $ g cm$^{-2}$, and a secondary black hole mass $  m_{\rm sBH} = 100 M_\odot  $. The orbital semi-major axis is derived from the period $  T_{\rm orb} = 63.7  $ ks and the central SMBH mass $  M_\bullet = 10^6 M_\odot  $:
\begin{equation}
a = \left( \frac{G M_\bullet T_{\rm orb}^2}{4\pi^2} \right)^{1/3} \approx 1.597 \, \mathrm{AU}.
\end{equation}
The Hill radius is then
\begin{equation}
R_{\rm H} = a \left( \frac{m_{\rm sBH}}{3 M_\bullet} \right)^{1/3} \approx 11.05 R_\odot.
\end{equation}
The relative velocity between the sBH and the disk gas is
\begin{equation}
v_{\rm rel} = 2 v_{\rm k} \sin\left( \frac{i}{2} \right) \propto a^{-1/2},
\end{equation}
where $  v_{\rm k} / c = 0.08  $ is the Keplerian speed. The Bondi radius is 
\begin{equation}
R_{\rm B} = \frac{2G m_{\rm sBH}}{v_{\rm rel}^2} \propto a.
\end{equation}
The effective interaction radius follows the empirical scaling in Eq.~(\ref{eq:R_eff}), 
\begin{equation}
R_{\rm eff} = 0.5 R_{\rm B}^{1/3} R_{\rm H}^{2/3} \propto a. \tag{\ref{eq:R_eff}}
\end{equation}
The residence time, i.e. the characteristic timescale over which the sBH traverses the high-density mid-plane of the disk, is defined as 
\begin{equation}
\tau \simeq \frac{R_{\rm eff}}{v_{\rm k} \sin i},
\end{equation}
approximated as the time for a ``ball" of size equal to the effective radius $  R_{\rm eff}  $ to cross the region of peak density. The definition does not depend on the disk scale height $H$, which is somewhat arbitrary and may vary between sources. Instead, we focus on the mid-plane where the gas density is highest and the gravitational focusing is most efficient. 
Assuming a Gaussian distribution, the ambient disk density is estimated with the mid-plane density as
\begin{equation}
\rho_0 = \frac{1}{\sqrt{2\pi}} \frac{\Sigma}{H}.
\end{equation}
The mass accumulation rate, treating the sBH as sweeping a spherical volume, is
\begin{equation}
\dot{M}_{\rm sBH} = \rho_0 \cdot \frac{4}{3} \pi R_{\rm eff}^2 v_{\rm rel}.
\end{equation}
The shocked mass over the residence time is then
\begin{equation}
M_{\rm sBH} = \dot{M}_{\rm sBH} \tau.
\end{equation}
The energy accumulation rate, reflecting gravitational heating as gas is pulled toward the sBH, is
\begin{equation}
\dot{E}_{\rm sBH} = \frac{G m_{\rm sBH}}{R_\odot^2} \rho_0 v_{\rm rel} \cdot \frac{4}{3} \pi R_{\rm eff}^3,
\end{equation}
where we adopt $  R_\odot^2  $ as a characteristic normalization scale to simplify the calculation of order-of-magnitude estimates of the gravitational pull effect. The choice is justified because the key physical length scales involved, such as the Bondi radius $R_{\rm B}$, the Hill radius $R_{\rm H}$, and the effective radius $R_{\rm eff}$ derived from simulations, all fall within a range from $\sim0.03\, R_\odot  $ to  $ \sim 10\, R_\odot  $ depending on the inclination and other parameters. By using a single representative distance $  R_\odot  $, we average the gravitational effect over this typical range, avoiding the need for a full integral over radius dependent contributions while still capturing the dominant behavior. 
We note that this choice is approximate. In the low-inclination regime, $R_{\rm eff}$ can exceed the assumed disk scale height, so the thin-disk and spherical sweeping assumptions become less accurate. However, the residence time $\tau$ is defined independently of $H$, which reduces sensitivity to this choice. A more self-consistent treatment is left for future work.

Thus, the total gravitational energy gain is
\begin{equation}
E_{\rm sBH} = \dot{E}_{\rm sBH} \tau.
\end{equation}
The expansion speed of the ejecta is estimated as the sound speed in the heated gas, assuming the energy is distributed over the mass:
\begin{equation}
c_s = \sqrt{\frac{E_{\rm sBH}}{M_{\rm sBH}}}.
\end{equation}
For low inclination ($  i = 18^\circ  $), $  v_{\rm rel} = 0.025c  $, $  R_{\rm B} = 0.68 R_\odot  $, $  R_{\rm eff} = 2.18 R_\odot  $, $  \tau = 204.6  $ s, $  \dot{M}_{\rm sBH} = 2.08 \times 10^{-7} M_\odot \, \mathrm{s}^{-1}  $, $  M_{\rm sBH} = 4.26 \times 10^{-5} M_\odot  $, $  \dot{E}_{\rm sBH} = 1.79 \times 10^{-11}  $ $  M_\odot R_\odot^2 \, \mathrm{s}^{-3}  $, $  E_{\rm sBH} = 3.52 \times 10^{46}  $ erg, and $  c_s \approx 0.022\,c  $.

For high inclination ($  i = 90^\circ  $), $  v_{\rm rel} = 0.113c  $, $  R_{\rm B} = 0.033 R_\odot  $, $  R_{\rm eff} = 0.80 R_\odot  $, $  \tau = 23.1  $ s, $  \dot{M}_{\rm sBH} = 1.26 \times 10^{-7} M_\odot \, \mathrm{s}^{-1}  $, $  M_{\rm sBH} = 2.91 \times 10^{-6} M_\odot  $, $  \dot{E}_{\rm sBH} = 3.96 \times 10^{-12}  $ $  M_\odot R_\odot^2 \, \mathrm{s}^{-3}  $, $  E_{\rm sBH} = 8.81 \times 10^{44}  $ erg, and $  c_s \approx 0.013\,c  $.

We note that the ejecta expansion speed scales with the secondary mass as $c_s \propto m_{\rm sBH}^{7/9}$. This quantitative relation explains why previous local simulations that adopted significantly more massive black holes (e.g., Lam et al. 2025, with $M_2 \sim 10^3$--$10^4\,M_\odot$) found ejecta velocities comparable to or exceeding the relative velocity. Increasing the black hole mass by one to two orders of magnitude naturally raises the ejecta velocity closer to $v_{\rm rel}$, consistent with those higher-mass results. In our stellar-mass black hole simulations, the much smaller interaction scale (especially at high inclination) leads to lower ejecta velocities, as expected from this scaling.

\begin{deluxetable*}{cccccccccc cc}
\tablewidth{0pt}
\tablecaption{Observed properties of the QPE sources and the corresponding derived quantities \label{tab:table2}}
\tablehead{
\colhead{Source} & \colhead{$t_{\mathrm{dur}}$(ks)} & \colhead{$t_{\mathrm{reccur}}$(ks)} & \colhead{$\log_{10}E_{\mathrm{QPE}}$ (erg)} & \colhead{$\log_{10}M_{\mathrm{SMBH}}$ ($M_{\odot}$)} & \colhead{a(AU)}    & \colhead{$R_{\mathrm{H}}(R_{\odot})$}& \colhead{$R_{\mathrm{B}}(R_{\odot})$}& \colhead{$i$ ($^\circ$)} & \colhead{$R_{\rm eff} (R_{\odot})$} 
& \colhead{$M_{\rm disk}$ ($M_\odot$)} & \colhead{Comment}
}
\startdata
GSN 069 & 4.50±0.60 & 29.90±9.50 & 46.11±0.11 & 6.28±0.72 &$1.90^{+2.07}_{-1.05}$&$10.60^{+2.14}_{-2.38}$&$0.34^{+9.58}_{-0.33}$& $20.2^{+34.5}_{-13.1}$ & $1.69^{+4.17}_{-1.21}$ & 1.3 & R \\
RX J1302 & 2.50±1.10 & 12.90±10.40 & 45.69±0.21 & 6.14±0.88 &$0.97^{+1.86}_{-0.81}$&$6.05^{+2.92}_{-4.03}$&$2.19^{+436.69}_{-1.68}$& $13.9^{+52.6}_{-12.7}$ & $1.33^{+15.08}_{-0.29}$ & 0.3 & R \\
eRO-QPE1 & 26.60±5.20 & 77.50±27.00 & 48.08±0.12 & 5.90±0.79 &$2.68^{+3.31}_{-1.58}$&$20.00^{+4.41}_{-4.97}$&$2.93^{+114.81}_{-2.84}$& $12.7^{+23.3}_{-8.6}$ & $5.27^{+15.35}_{-3.93}$ & 2.5 & R \\
eRO-QPE2 & 1.70±0.10 & 8.30±0.80 & 45.77±0.04 & 5.43±0.79 &$0.42^{+0.40}_{-0.21}$&$4.51^{+0.29}_{-0.30}$&$1.06^{+15.86}_{-0.99}$& $14.4^{+16.1}_{-7.6}$ & $1.39^{+2.26}_{-0.86}$ & 0.1 & R \\
eRO-QPE3 & 8.40±0.45 & 72.00±7.10 & 45.83±0.13 & 5.53±0.79 &$1.92^{+1.83}_{-0.94}$&$19.04^{+1.23}_{-1.27}$&$0.17^{+0.99}_{-0.17}$& $115.8^{+64.2}_{-64.2}$ & $1.96^{+1.94}_{-1.94}$ & 1.3 & R \\
eRO-QPE4 & 3.60±0.40 & 50.80±10.10 & 46.46±0.21 & 7.31±0.75 &$5.96^{+6.00}_{-3.07}$&$15.09^{+1.94}_{-2.07}$&$0.32^{+7.65}_{-0.30}$& $11.4^{+18.4}_{-7.2}$ & $2.08^{+4.53}_{-1.42}$ & 12.6 & H \\
eRO-QPE5 & 51.84±9.50 & 319.68±1.73 & 47.53±0.09 & 7.45±0.52 &$22.62^{+11.22}_{-7.50}$&$51.44^{+0.19}_{-0.19}$&$0.17^{+0.83}_{-0.14}$& $26.0^{+17.6}_{-10.1}$ & $3.82^{+3.11}_{-1.73}$ & 180.9 & E \\
AT2019qiz & 31.90±1.60 & 175.80±19.40 & 47.68±0.04 & 6.27±0.76 &$6.14^{+5.66}_{-2.97}$&$34.52^{+2.50}_{-2.59}$&$0.48^{+6.90}_{-0.45}$& $31.4^{+41.8}_{-16.6}$ & $4.15^{+6.66}_{-2.61}$ & 13.3 & H \\
ZTF19acnskyy & 125.50±14.10 & 440.90±85.20 & 47.99±0.17 & 6.34±0.66 &$11.95^{+10.37}_{-5.71}$&$63.73^{+7.97}_{-8.50}$&$0.22^{+4.33}_{-0.22}$& $61.8^{+118.2}_{-38.1}$ & $4.83^{+9.47}_{-4.81}$ & 50.5 & E \\
AT2022upj & 59.00±8.60 & 172.20±54.20 & 47.77±0.12 & 6.38±0.56 &$6.59^{+5.56}_{-3.26}$&$34.05^{+6.82}_{-7.58}$&$0.58^{+9.82}_{-0.55}$& $25.9^{+38.0}_{-10.3}$ & $4.39^{+8.56}_{-2.95}$ & 15.4 & H \\
J2049 & 1.20±0.05 & 9.50±1.00 & 45.36±0.12 & 5.00±0.50 &$0.33^{+0.19}_{-0.12}$&$4.94^{+0.34}_{-0.35}$&$0.43^{+2.89}_{-0.37}$& $33.4^{+32.5}_{-15.5}$ & $1.09^{+1.17}_{-0.58}$ & 0.0 & R \\
AT2019vcb & 54±18 & 144±108 & 48.81±0.15 & 6.81±0.13 &$8.13^{+4.92}_{-5.21}$&$30.22^{+13.67}_{-18.23}$&$4.53^{+199.57}_{-4.23}$& $6.2^{+7.6}_{-1.2}$ & $8.03^{+28.6}_{-6.27}$ & 23.4 & E \\
\enddata
\tablecomments{
Using the source parameters compiled in \cite{goodwin2025PASA...42..130G}, we computed the orbital inclination and effective radius for each source using the semi-analytic model described above, assuming a disk surface density of $10^{6}\,\mathrm{g\,cm^{-2}}$ and a sBH mass of $100\,M_{\odot}$. 
Disk mass is calculated as $M_{\rm disk} = \pi a^2 \Sigma$ with the fiducial surface density $\Sigma = 10^6$ g cm$^{-2}$ (central value of $a$ is used). 
Comment: R = reasonable ($\lesssim 5\,M_\odot$), H = high ($5$--$20\,M_\odot$), E = extremely high ($>20\,M_\odot$). Sources labeled 'H' and 'E' require either a significantly lower disk surface density, a lower orbital inclination, or a more massive secondary black hole to yield physically plausible disk masses.
}
\end{deluxetable*}

This simple semi-analytical framework qualitatively reproduces the simulation trend of higher accreted mass (one order of magnitude), faster speed of ejecta (a factor of two) and energy (two orders of magnitude) at low inclinations, driven by larger $  R_{\rm eff}  $ (from slower $  v_{\rm rel}  $) and longer residence time $  \tau  $. Besides, the $  R_{\rm eff}^3  $ scaling in $  \dot{E}_{\rm sBH}  $ emphasizes volume-like heating, amplifying the energy difference. Modulations induced by these parameters match observed QPE variations, with low inclinations causing strong flares and high inclinations leading to weaker events, explaining alternating patterns and diversity in recurrence and luminosities.

In the semi-analytical framework, the residence time scales linearly with the orbital period. Because $  R_{\rm eff} \propto a  $ while the vertical crossing velocity scales as $  v_k \sin i \propto a^{-1/2}  $, we obtain $  \tau \propto a^{3/2}  $. Since the orbital period satisfies $  T_{\rm orb} \propto a^{3/2}  $, it follows that $  \tau \propto T_{\rm orb}  $. This naturally explains a linear correlation between flare duration and recurrence time, consistent with observations \citep{arcodia2025ApJ...989...13A, baldini2026A&A...706L..15B}.

Recent observational studies have provided initial constraints on the fraction of TDEs that produce QPEs. Using a volumetric rate analysis, \citet{2024A&A...684L..14A} estimated that the QPE formation rate is significantly lower than the TDE rate, implying that only a small fraction (on the order of $\sim$1\% or less, depending on the assumed QPE lifetime) of TDEs may develop observable QPE activity. In contrast, \citet{chakraborty2025ApJ...983L..39C} performed a statistical analysis of optically selected TDEs and found that roughly 9\% develop QPEs within a few years. 

Using the semi-analytic model developed in this work, we estimate the orbital inclinations required to reproduce the observed QPE energies for the sources compiled by \cite{goodwin2025PASA...42..130G}. As listed in Table \ref{tab:table2}, most sources are consistent with relatively low orbital inclinations ($\lesssim 30^\circ$--$40^\circ$, or $\lesssim \pi/3$) under the fiducial assumptions. A few sources (particularly eRO-QPE5, ZTF19acnskyy, and AT2019vcb) imply extremely high disk masses ($>20\,M_\odot$), which are difficult to reconcile with a TDE origin. Although these extreme cases can be accommodated by adopting a lower disk surface density or a more massive secondary black hole while still permitting relatively low inclinations, they remain challenging for both the star-disk and sBH-disk models and warrant dedicated follow-up studies to clarify their physical origin.

Nevertheless, the majority of known QPE sources still favor relatively low orbital inclinations, implying that only a modest fraction of randomly oriented EMRIs can produce observable QPEs. For randomly oriented orbits, the corresponding solid-angle fraction is
\begin{equation}
P(i < \pi/3) = \frac{1-\cos(\pi/6)}{2}  \approx 6.7\%.
\end{equation}
This value represents an upper limit, since our estimate is purely geometric and assumes that every supermassive black hole in host galaxies that have recently gone through an AGN activity is accompanied by a stellar-mass black hole companion. AGN disks can both capture existing stellar-mass black holes via gas drag and generate new ones through in-situ star formation in their outer regions; the newly formed black holes can subsequently migrate inward and contribute to the population available for later TDE-disk impacts \citep{Jiang:2025jbd}. In reality, the actual fraction of systems hosting such EMRIs is likely lower, which would bring the predicted occurrence rate into even better agreement with current observational estimates. This demonstrates that the preference for relatively low orbital inclinations in the sBH-disk model naturally provides a geometric selection effect that is broadly consistent with the observed occurrence rate of QPEs among TDEs.

\subsection{Implications}

Based on our semi-anlytical model, fitting the energy budget of GSN 069 does not necessarily require a secondary as massive as $100 M_\odot$. For a disk surface density of $10^{6}\,\mathrm{g\,cm^{-2}}$, an sBH with a mass of $\sim 50\,M_\odot$ can already provide sufficient energy at an inclination of $i = \pi/10$, and it is expected that the required mass increases at higher inclinations. 
Since the burst energy of different QPEs span $\sim10^{44}$--$10^{48}\,\mathrm{erg}$, the sBH-disk model can accommodate a variety of cases within this range, providing a viable explanation. We note that converting the deposited kinetic energy into observable X-ray radiation involves radiative efficiency and spectral modeling, which are not addressed in the current work and remain important uncertainties for future study.

Because the EMRI usually has a mild eccentricity, the two impacts in one orbital period may not necessarily have the same inclination angle, Such a discrepancy may contribute to the strong-weak patterns of outbursts. In this work, however, we fixed the inclination to investigate the origins of ejecta mass and energy source for sBHs impacting disks with a realistic global potential. Effects caused by variations of consecutive impacts are left for future studies. 

In addition, we refrain from studying BHs more massive than $\sim 100\,M_\odot$. 
For more massive secondary black holes, the larger effective interaction radius and Hill radius during disk crossings lead to stronger perturbations, and repeated passages can progressively disrupt the accretion disk's structure. Once the disk can no longer maintain its morphology, it becomes unable to support the stable periodicity required in QPE models. Therefore, some QPEs with poorer periodic coherence (e.g., eRO-QPE1) may be the result of interactions involving more massive black holes.
In future work, we will further investigate the secular evolution of the system, encompassing both the disk structure and orbital dynamics, under perturbations from a binary BH system.
This mechanism may be related to the newly discovered QPE source Ansky \citep{ansky2025NatAs...9..895H, ansky2025arXiv250916304H}--- whose long-term monitoring shows a doubling of the recurrence period, and more generally to sources with irregular recurrence and sources whose recurrence period evolves secularly over time. 

Interestingly, radiation-transport calculations by \citet{vurm2025ApJ} indicate that the star-disk collision model prefers high-inclination orbits (high relative velocity) to reach the photon-starved regime capable of producing bright soft X-ray flares. In contrast, our sBH-disk simulations show that low-inclination collisions are actually more favorable: the larger effective interaction radius allows substantially more disk gas to be gravitationally focused and heated, leading to higher characteristic ejecta velocities and energy budgets despite the lower impact speed. This opposite inclination dependence may help to differentiate the two models.

Note that numerical simulations of extremely long-period QPEs or very thick accretion disks remain challenging, and our current study does not include radiation transfer. Addressing these aspects, together with multi-wavelength observations and more refined simulations, will be essential to identify sBH-disk collisions and to further improve our understanding of QPE formation mechanisms.

\section{Summary} \label{sec:summary}
With a series of 3D global mesh-free hydrodynamic simulations, we carefully compare different EMRI+disk models taking into account the potential of the central SMBH, which is normally ignored. The star-disk and sBH-disk collisions exhibit different inclination dependence in both the energy output and the diffusion timescale. 
This difference arises from the distinct mechanisms of energy generation: in the star-disk model, energy is primarily transferred through direct collision, accelerating the mass within the stellar cross-section, whereas in the sBH-disk model, energy is imparted via gravitational pull on the disk gas, with the effective interaction radius $  R_{\rm eff}  $ playing a key role in scaling the shocked mass and energy. 

Although sufficient energy can be injected into the disk by stellar collisions, the resulting eruptions are highly asymmetric. In other words, either only one side of eruptions could be observed, or consecutive eruptions exhibit large variations in brightness. Neither of these features are consistent with the observations. In addition, durations of outbursts produced by stellar impacts are in a relatively narrow range, making it difficult to account for the large diversity of QPEs, especially for the ultra-long-duration events.
Besides, the sustainability of a main sequence star against tides in an orbit with such short periods around the SMBH is another concern.

In contrast, the sBH-disk model is more attractive. A stellar-mass black hole $  m_{\rm sBH} \sim 50 M_\odot  $ passing through the accretion disk at a low inclination can produce energetic nearly symmetric two-sided ejecta, sufficient to explain sources like GSN 069. The moderate strong-weak alternation between consecutive flares of QPEs can be attributed to variations between two impacts in one orbital period of the sBH. 
Our semi-analytical model, calibrated to simulations, shows that sBH impacts can generate significantly higher energy budgets than previous Bondi-only estimates, making sBHs viable for powering a wide range of observed QPE energies without requiring IMBHs. Incorporating the effective radius dependence $  R_{\rm eff} \simeq 0.5 R_{\rm B}^{1/3} R_{\rm H}^{2/3}  $ corrects this, showing that the Hill radius dominates the scaling and enables realistic energy generation even for sBHs. 
Besides, the range of eruption durations is more consistent with the observations. Unlike stellar EMRIs the lifetime of an sBH is not a limiting factor. However, it is necessary to consider the long-term stability of the disk under the perturbation of an sBH. It can be straightforward to demonstrate that the disk  remains stable over thousands of passages of an sBH with global $N$-body simulations \citep{2012A&A...537A.128R} containing the SMBH and sBH treating the disk with test particles.

Furthermore, the sBH-disk model naturally predicts a relatively low occurrence rate of QPEs among TDEs. Reproducing the observed flare energies generally requires relatively low orbital inclinations, so only a modest fraction ($\sim$ 6.7\%) of randomly oriented EMRIs can produce bright QPEs. This geometric fraction should be viewed as an upper limit, since our estimate assumes that every supermassive black hole in host galaxies that have recently experienced AGN activity is accompanied by a stellar-mass black hole that is captured or formed in AGN disks. Combined with realistic disk mass constraints, this helps explain why QPEs are observed in only a small fraction of TDEs.

In summary, we conclude that the dynamics of the sBH-disk model matches observations more closely, providing a reasonable and robust explanation for the origin of QPEs. Future work will further explore the parameter space and incorporate radiative transfer and relativistic corrections. We plan to perform systematic, source-by-source modeling of cataloged QPEs to enable quantitative comparison with observations. High-cadence, multi-wavelength observations, together with gravitational wave detections, will be essential to test the sBH-disk model and constrain parameters of QPE systems. In particular, confirming that the EMRIs responsible for QPEs are sBHs rather than stars would be highly significant. It would offer a unique opportunity to probe the environment around supermassive black holes, implying that such sBHs may be ubiquitous and positioning them as promising targets for space-based gravitational wave observatories \citep{lui2025arXiv250807961L}.

\begin{acknowledgments}
We are grateful to Junping Chen, Xian Chen, Sierra Dodd, Alessia Franchini, Wenyuan Guo, Xiaoshan Huang, Ning Jiang, Shuo Li, Bo Ma, Erlin Qiao, Enrico Ramirez-Ruiz, Ruiqi Yang, Bei You, Yong Zhang and Cong Zhou for valuable exchanges. In addition, we thank the participants of the TDE FORUM (Full-process Orbital to Radiative Unified Modeling) online seminar series for their inspiring discussions. We also thank the anonymous referee for thoughtful suggestions which resulted in a greatly improved paper. This work is supported by the National Natural Science Foundation of China (grant No. 42530203) and the Guangdong Basic and Applied Basic Research Foundation (grant No. 2021B1515020090). H.D. acknowledge support from the National Science Foundation of China via grant No. 12522307.
\end{acknowledgments}

\bibliography{ms}{}
\bibliographystyle{aasjournalv7}

\end{CJK*}
\end{document}